\documentclass[11pt,twocolumn]{article}
\usepackage{graphicx}
\usepackage[numbers]{natbib}
\usepackage{xcolor}
\usepackage{hyperref}
\usepackage[T1]{fontenc}
\usepackage{lmodern}
\usepackage{setspace}
\usepackage{fancyhdr}
\usepackage[margin=1.8cm]{geometry}
\fancypagestyle{plain}{%
	\fancyhf{}
	\fancyhead[L]{ML for RNA Secondary Structure Prediction}
	\fancyhead[R]{Sacco \thepage}
}
\title{Machine Learning for RNA Secondary Structure Prediction: a review of current methods and challenges\thanks{Updated version of \citet{saccoMachineLearningRNA}. This version revises the Foundation Models section to reflect rapid developments since publication; all other sections are unchanged apart from formatting.}}
\author{Giuseppe Sacco, Giovanni Bussi, Guido Sanguinetti\thanks{Corresponding author: \texttt{gsanguin@sissa.it}}\\
\small All authors: Scuola Internazionale Superiore di Studi Avanzati, SISSA,\\ \small Trieste, Italy, 34136}
\date{\today}

\begin{document}

\twocolumn[
\maketitle

\noindent\textbf{Running head title:} ML for RNA Secondary Structure Prediction


\begin{abstract}
Predicting the secondary structure of RNA is a core challenge in computational biology, essential for understanding molecular function and designing novel therapeutics. The field has evolved from foundational but accuracy-limited thermodynamic approaches to a new data-driven paradigm dominated by machine learning and deep learning. These models learn folding patterns directly from data, leading to significant performance gains. This review surveys the modern landscape of these methods, covering single-sequence, evolutionary-based, and hybrid models that blend machine learning with biophysics. A central theme is the field's "generalization crisis," where powerful models were found to fail on new RNA families, prompting a community-wide shift to stricter, homology-aware benchmarking. In response to the underlying challenge of data scarcity, RNA foundation models have emerged, learning from massive, unlabeled sequence corpora to improve generalization. Finally, we look ahead to the next set of major hurdles---including the accurate prediction of complex motifs like pseudoknots, scaling to kilobase-length transcripts, incorporating the chemical diversity of modified nucleotides, and shifting the prediction target from static structures to the dynamic ensembles that better capture biological function. We also highlight the need for a standardized, prospective benchmarking system to ensure unbiased validation and accelerate progress.
\end{abstract}

\vspace{0.5em}
\noindent\textit{\small Note: this is an updated version of the article published as \citet{saccoMachineLearningRNA}. The Foundation Models section has been expanded to reflect rapid progress in the field since the journal version was accepted; the remainder of the manuscript is unchanged apart from formatting.}
]

\clearpage

\section{Introduction}

\subsection{The Expanding World of RNA}
Ribonucleic acid (RNA) molecules are fundamental biomolecules that fulfill a wide range of biological functions, extending far beyond their traditional role as genetic information carriers from DNA to proteins \citep{doudnaChemicalRepertoireNatural2002, morrisRiseRegulatoryRNA2014}.
In recent years, the diverse functionalities of non-coding RNAs (ncRNAs)—RNA molecules that are not translated into proteins—have been increasingly recognized, impacting processes such as development, cell differentiation, and disease \citep{statelloGeneRegulationLong2021}.
The sheer variety of RNA species continues to expand, with many cataloged in comprehensive databases like Rfam and RNAcentral \citep{kalvariRfam14Expanded2021, thernacentralconsortiumRNAcentralHubInformation2019}.
Long non-coding RNAs (lncRNAs), in particular, are gaining attention for their critical roles in various intracellular regulatory processes in eukaryotes, including humans \citep{mattickNoncodingRNAsArchitects2001, koppFunctionalClassificationExperimental2018}.
Their association with diseases like cancer and neurodegenerative disorders has positioned them as promising new targets for drug discovery \citep{pengLncRNAmediatedRegulationCell2017, wuRolesLongNoncoding2013}.

\subsection{The Centrality of Secondary Structure}
The functional capabilities of RNA molecules are inextricably linked to their intricate structures \citep{brionHIERARCHYDYNAMICSRNA1997}.
RNA structure is typically described through a hierarchical model, beginning with the primary structure, which is the one-dimensional sequence of nucleotides (Adenine, Cytosine, Guanine, and Uracil).
The secondary structure then emerges from the primary sequence through the formation of hydrogen bonds between complementary bases.
These interactions primarily involve Watson-Crick base pairs (A-U and G-C) and wobble base pairs (G-U), which are the most common in RNA secondary structures \citep{tinocoHowRNAFolds1999}.
This secondary structure forms rapidly from the primary sequence, accompanied by a significant loss of energy, and critically serves as a foundational scaffold that guides the subsequent folding of the RNA molecule into its complex three-dimensional (tertiary) structure \citep{brionHIERARCHYDYNAMICSRNA1997}.
Consequently, a thorough understanding of RNA secondary structure is paramount for deciphering RNA functions, developing RNA-based therapeutics, and accurately predicting the molecule's final three-dimensional conformation.
RNA secondary structures are known to be evolutionarily conserved among RNA species \citep{eddyRNASequenceAnalysis1994}.
RNA secondary structures can be formally represented in several ways.
One common method is using binary-valued triangular matrices, where a value of $y_{ij} = 1$ indicates that bases at positions $i$ and $j$ are paired.
Another widely used representation is the dot-bracket notation.
In its simplest form, this notation uses matching parentheses `(` and `)` to denote paired bases and dots `.` for unpaired bases.
This convention is sufficient for representing secondary structures with entirely nested base pairs, which are known as pseudoknot-free structures.
However, many biologically important RNAs contain pseudoknots, which are structural motifs characterized by non-nested base pairs where bases within a loop form pairs with bases outside of that loop \citep{bellaousovProbKnotFastPrediction2010}.
Because these crossing interactions cannot be described with a single type of parenthesis, the dot-bracket notation is extended to include additional bracket types (e.g., `[` and `]`, or `\{` and `\}`) to represent these more complex topologies.

\subsection{Experimental Limits and the Sequence-Structure Gap}
Despite the critical importance of RNA structure, its experimental determination, particularly for secondary and tertiary structures, remains a slow, costly, and technically demanding endeavor \citep{tinocoHowRNAFolds1999, holbrookStructuralPrinciplesLarge2008, cruzDynamicLandscapesRNA2009, strobelHighthroughputDeterminationRNA2018}.
High-resolution methods like X-ray crystallography, nuclear magnetic resonance (NMR), and cryogenic electron microscopy (cryo-EM), while powerful, inherently suffer from low throughput.
This means that only a minute fraction of the vast number of known RNAs have had their structures experimentally elucidated \citep{tinocoHowRNAFolds1999, holbrookStructuralPrinciplesLarge2008, cruzDynamicLandscapesRNA2009}.
Furthermore, even these gold-standard techniques often provide data that is ensemble- or time-averaged.
Since RNA molecules can exist as a heterogeneous ensemble of conformations, this averaging can obscure the presence of less populated but functionally important alternative structures \citep{cruzDynamicLandscapesRNA2009}.
Consequently, computational methods are often required to deconvolute this experimental information and model the full structural ensemble.
Chemical probing techniques like Selective 2'-Hydroxyl Acylation Analyzed by Primer Extension (SHAPE) \citep{merinoRNAStructureAnalysis2005} and Dimethyl Sulphate (DMS) \citep{peattieChemicalProbesHigherorder1980} are routinely used to improve the accuracy of predictions
based on thermodynamic models
\citep{weeksAdvancesRNAStructure2010, mustoePervasiveRegulatoryFunctions2018, strobelHighthroughputDeterminationRNA2018}.
However, the information content of chemical probing experiments is limited
\citep{sukosdEvaluatingAccuracySHAPEdirected2013}, making the determination of the structure of large RNA molecules
still very difficult.
This, together with the decreasing cost of sequencing technologies, has led to a significant "sequence-structure gap": an enormous volume of RNA sequence data is continuously generated, yet the number of experimentally determined RNA structures remains severely limited.
For instance, less than 0.01\% of the millions of non-coding RNAs listed in RNAcentral have experimentally validated structures \citep{strobelHighthroughputDeterminationRNA2018}.
This substantial gap underscores the urgent and persistent need for accurate, cost-effective, and high-throughput computational prediction methods.

\subsection{A Brief History of Computational Methods}
The computational prediction of RNA secondary structure has been a prominent area of research since the 1970s \citep{zukerOptimalComputerFolding1981}.
Historically, the field was dominated by \textbf{thermodynamics-based methods}.
These approaches describes the energy of a folded RNA using a nearest-neighbor energy model \citep{tinocoImprovedEstimationSecondary1973}
and then identify either the minimum free-energy (MFE) structure or the entire partition function
using dynamic programming
\citep{nussinovAlgorithmsLoopMatchings1978,zukerMfoldWebServer2003, mathewsExpandedSequenceDependence1999, hofackerFastFoldingComparison1994}.
However, the performance of these methods eventually plateaued due to fundamental limitations of the nearest-neighbor model and further simplifying assumptions that inherently precluded the prediction of complex but biologically important features such as pseudoknots and tertiary contacts.
Moreover, these approaches depend on a fixed catalog of energy parameters obtained through labor-intensive experiments.
While the thermodynamic paradigm was central, other classical approaches also made important contributions.
\textbf{Co-evolutionary methods} leveraged multiple sequence alignments (MSAs) to identify conserved base pairs through correlated mutations \citep{eddyRNASequenceAnalysis1994, nawrockiInfernal11100fold2013}.
While powerful when applicable, this approach is fundamentally constrained by the "homology bottleneck": it requires a deep and diverse MSA to distinguish signal from noise,
but constructing a meaningful MSA often required prior structural information.
Furthermore, this approach is completely inapplicable to the vast number of "orphan" RNAs for which no homologs are known.
\textbf{Stochastic Context-Free Grammars (SCFGs)} provided a formal probabilistic framework for modeling RNA structure \citep{sakakibaraStochasticContextfreeGrammers1994, durbinBiologicalSequenceAnalysis1998}, but, like their thermodynamic counterparts, standard implementations were typically restricted to pseudoknot-free structures and struggled to capture the full complexity of RNA folding without becoming computationally intractable.
The limitations inherent in these classical paradigms created a clear need for new approaches.
The first wave of \textbf{Machine Learning (ML)} emerged as a direct response to the shortcomings of the thermodynamic model.
These methods sought to replace the fixed experimental energy parameters with richer, data-driven scoring functions, while still relying on the classical dynamic programming machinery \citep{doCONTRAfoldRNASecondary2006, andronescuComputationalApproachesRNA2010}.
More recently, the field has been revolutionized by rapid advancements in \textbf{Deep Learning (DL)} technologies and the increasing availability of large-scale RNA datasets.
These methods represent a paradigm shift, moving from explicit physical or evolutionary models to learning the complex sequence-to-structure mapping directly from data.
While many prominent DL approaches are designed to be end-to-end, another successful branch of research uses deep learning to create hybrid models that enhance classical frameworks \citep{singhImprovedRNASecondary2021, fuUFoldFastAccurate2022}.
This data-driven revolution has led to notable improvements in prediction accuracy \citep{huangLinearFoldLineartimeApproximate2019, wangRNADiffFoldGenerativeRNA2025}.
However, it has also introduced new challenges; these data-hungry models are often susceptible to overfitting and can struggle to generalize to novel RNA families \citep{szikszaiDeepLearningModels2022}, a critical limitation that remains a central focus of current research.

\subsection{Review Scope}

This review surveys computational methods for RNA secondary structure prediction across three pillars: classical baselines, data and generalization, and modern deep learning.
We first recap the thermodynamics-, evolutionary-, and grammar-based foundations to establish assumptions, strengths, and long-standing limitations that motivate data-driven approaches.
We then examine datasets, curation pitfalls, and homology-aware evaluation, highlighting the field's generalization crisis and the emerging norms for rigorous benchmarking.
The core of the review classifies deep learning methods by input regime—single-sequence (ab initio), evolutionary (MSA-based), and biophysical hybrids—and synthesizes trends such as thermodynamic integration and end-to-end predictors.
We discuss the emerging frontier of RNA foundation models and their potential to mitigate data scarcity.
Finally, we map persistent challenges and opportunities: accurate modeling of pseudoknots and non-canonical pairs, scaling to kilobase-length RNAs, representing chemical modifications and environmental context, and shifting targets from single MFE structures to dynamic ensembles that better capture the dynamic nature of RNA.
We conclude with a forward-looking discussion on standards, including prospective community benchmarks to enable fair comparisons and sustained progress.

\section{Classical Methods}

\begin{figure*}[t]
\centering
\includegraphics[width=\linewidth]{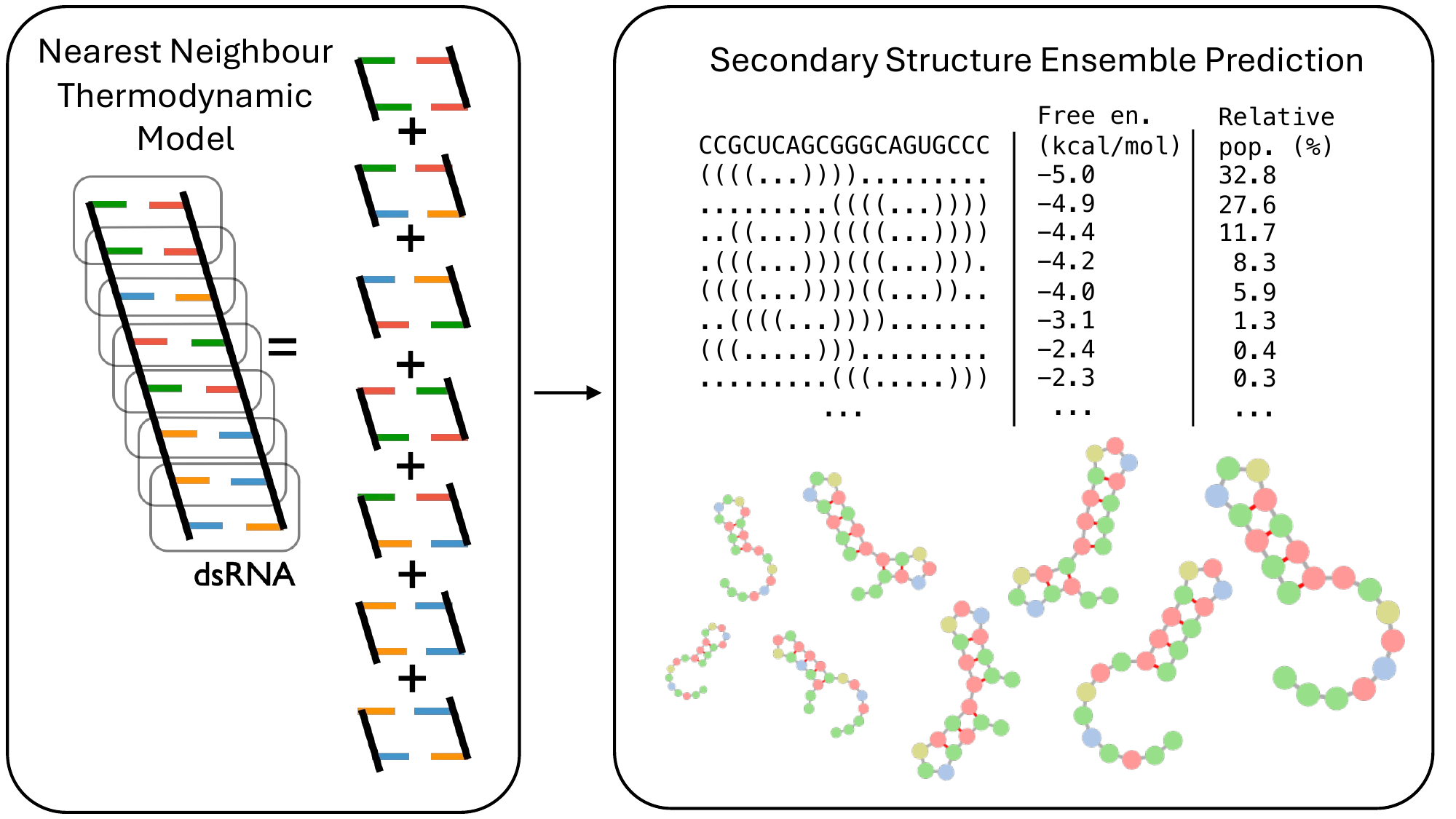}
\caption{Schematic representation of thermodynamics-based RNA secondary structure prediction.
The free energy of a structure is computed with the Nearest Neighbor model (left panel) as the sum of contributions from individual structural elements, enabling efficient dynamic programming algorithms to enumerate and predict the relative population of all of the possible secondary structures for a given RNA sequence (right panel).
Secondary structure visualization generated with Forna \citep{kerpedjievFornaForcedirectedRNA2015}.
}
\label{fig:thermodynamic_models}
\end{figure*}
The computational prediction of RNA secondary structure has a long history, with methods that can be broadly categorized into four classical paradigms: thermodynamics-based, co-evolutionary, grammar-based, and early machine learning approaches.

\subsection{Thermodynamics-Based Methods}
The earliest and most dominant approach to RNA secondary structure prediction is based on the principles of thermodynamics, as schematically illustrated in Figure~\ref{fig:thermodynamic_models}.
This model posits that the most stable RNA secondary structure is the one possessing the minimum free energy (MFE) \citep{tinocoImprovedEstimationSecondary1973, brionHIERARCHYDYNAMICSRNA1997, holbrookRNAStructureLong2005}.
Algorithms based on dynamic programming (DP), such as the Zuker-Stiegler algorithm \citep{zukerOptimalComputerFolding1981}, are designed to efficiently search for these optimal structures \citep{nussinovAlgorithmsLoopMatchings1978,hofackerFastFoldingComparison1994}.
This foundational approach is implemented in widely used software packages such as \textbf{Mfold} \citep{zukerMfoldWebServer2003} and its successor \textbf{UNAFold} \citep{markhamUNAFold2008}, which compute the MFE structure and can also generate suboptimal foldings.
The \textbf{ViennaRNA Package} \citep{lorenzViennaRNAPackage202011}, containing the key component \textbf{RNAfold}, implements a similar DP approach and is a benchmark tool in the field.
These tools iteratively build optimal structures for subsequences in polynomial time, typically achieving a computational complexity of $O(L^3)$ for an RNA sequence of length $L$, which can be slow for long sequences.
A significant limitation of these approaches is their heavy reliance on a fixed set of experimentally determined energy parameters (e.g., Turner’s rules) \citep{turnerNNDBNearestNeighbor2010}.
ViennaRNA and other softwares like \textbf{RNAstructure} \citep{reuterRNAstructureSoftwareRNA2010} are designed to seamlessly incorporate constraints from chemical probing experiments, which can significantly improve accuracy.
A critical shortcoming of most traditional DP-based algorithms is their inherent restriction to predicting "nested" structures, meaning they cannot model pseudoknots.
Pseudoknots are non-nested base-pair interactions that are biologically significant, occurring in roughly 40\% of all RNAs \citep{staplePseudoknotsRNAStructures2005, holbrookStructuralPrinciplesLarge2008}.
RNAstructure, for instance, includes \textbf{ProbKnot} \citep{bellaousovProbKnotFastPrediction2010}, a Maximum Expected Accuracy (MEA) method that predicts the presence of pseudoknots, and \textbf{ShapeKnots} \citep{hajdinAccurateSHAPEdirectedRNA2013}, which uses SHAPE data to guide pseudoknot prediction.
The general problem of predicting the lowest free-energy structures that include arbitrary pseudoknots has been proven to be NP-complete or NP-hard, making it computationally prohibitive for exact energy-based methods.
In addition, it is difficult to know by the sole secondary structure if a given pseudoknot would be achievable in a three-dimensional model, with heuristics that have been proposed to address this issue \citep{bonTT2NENovelAlgorithm2011}.

\subsection{Stochastic Context-Free Grammars (SCFGs)}
This foundational concept was established by Eddy \& Durbin \citep{eddyRNASequenceAnalysis1994}, who introduced covariance models (CMs) as a way to describe both the sequence and structure consensus of an RNA family.
The software \textbf{Infernal} \citep{nawrockiInfernal11100fold2013} is the engine that builds and uses these CMs, most famously to maintain the Rfam database \citep{kalvariRfam14Expanded2021}.
Prediction tools like \textbf{Pfold} \citep{knudsenPfoldRNASecondary2003} and \textbf{EvoFold} \citep{pedersenIdentificationClassificationConserved2006} utilize advanced SCFGs that incorporate explicit evolutionary models to predict a consensus structure from an alignment.

\subsection{Co-evolutionary Methods}
This paradigm leverages the principle that RNA secondary structures are often conserved across evolution, even when the primary sequence diverges.
The "alignment folding" strategy is operationalized by tools like \textbf{RNAalifold} \citep{bernhartRNAalifoldImprovedConsensus2008}, part of the ViennaRNA suite. It computes a consensus structure by combining an averaged thermodynamic energy term with an additional score for covariation, the canonical signal of which is a compensatory mutation (e.g., a G-C pair mutating to an A-U pair).
The align-then-fold approach relies heavily on the quality of the Multiple Sequence Alignment (MSA) used as input \citep{gardnerComprehensiveComparisonComparative2004}.
This reliance is addressed by methods that utilize the Covariance Models (CMs), originally defined in the Stochastic Context-Free Grammar (SCFG) paradigm \citep{eddyRNASequenceAnalysis1994, rivasRangeComplexProbabilistic2012, satoRecentTrendsRNA2023}.
The software Infernal \citep{nawrockiInfernal11100fold2013} is central to this, using CMs built from initial alignments or consensus structures \citep{nawrockiInfernal11100fold2013, chenMARSRNAcmap3Master2024} to perform highly sensitive homology searches against sequence databases \citep{nawrockiInfernal11100fold2013, zhangRNAcmapFullyAutomatic2021} and generate refined, structure-aware MSAs via its cmalign program \citep{chenMARSRNAcmap3Master2024, nawrockiInfernal11100fold2013}.
Because CMs explicitly model both sequence and secondary structure conservation \citep{eddyRNASequenceAnalysis1994, nawrockiInfernal11100fold2013, zhangMultipleSequenceAlignmentbased2024}, the resulting alignments provide superior input quality for subsequent structural inference methods, including those based on Direct Coupling Analysis (DCA) \citep{cuturelloAssessingAccuracyDirectcoupling2020,zhangMultipleSequenceAlignmentbased2024,pucciEvaluatingDCAbasedMethod2020,deleonardisDirectCouplingAnalysisNucleotide2015}.
The "simultaneous" strategy is based on the Sankoff algorithm \citep{sankoffSimultaneousSolutionRNA2006}, a DP method that simultaneously aligns sequences and infers a consensus structure.
However, its full implementation is computationally intractable. Practical tools are therefore restricted versions; \textbf{Dynalign} \citep{mathewsDynalignAlgorithmFinding2002} implements this for two sequences using a full thermodynamic model and is effective for divergent sequences, while \textbf{Foldalign} \citep{sundfeldFoldalign25Multithreaded2016} uses a simpler scoring scheme, often for finding short, conserved local motifs.

\subsection{Early Machine Learning Approaches}
To overcome the inherent limitations of purely thermodynamics-based methods, early machine learning (ML) approaches were introduced to RNA secondary structure prediction.
This data-driven approach allowed for the development of more accurate models by enabling a richer and more comprehensive parameterization than what was feasible through wet-lab experiments alone \citep{satoRecentTrendsRNA2023}.
This era saw a variety of ML techniques applied to learn better scoring functions. \textbf{SimFold} \citep{andronescuEfficientParameterEstimation2007}, for example, used a regularized linear model to optimize Turner's energy parameters to better fit training data. \textbf{TORNADO} \citep{rivasRangeComplexProbabilistic2012}, a flexible framework for exploring complex Stochastic Context-Free Grammars, used Maximum Likelihood training to parameterize its probabilistic models.
However, the most influential methods were based on discriminative training.
A landmark example is \textbf{CONTRAfold} \citep{doCONTRAfoldRNASecondary2006}, which utilized conditional log-linear models (CLLMs).
Instead of learning a generative model of the joint probability of a sequence and structure like an SCFG, CONTRAfold directly models the conditional probability of a structure given a sequence.
Its approach releases the strict constraints of a formal grammar, allowing for a more flexible, feature-rich scoring system.
The model learns weights for these features from data, and the resulting scores are then optimized using a DP algorithm analogous to the classic Zuker algorithm.
A critical evolutionary step was taken by \textbf{ContextFold} \citep{zakovRichParameterizationImproves2011}, which provided a powerful proof-of-concept for "rich parameterization."
It demonstrated that the field's performance plateau could be broken by abandoning the constraint of a small, physically-derived parameter set.
By using a discriminative online learning algorithm, ContextFold effectively trained a model with approximately 70,000 parameters describing fine-grained structural and sequential contexts, leading to a nearly 50\% reduction in prediction error over the then-state-of-the-art.
Here, sequential context denotes the identities and positions of nucleotides at fixed offsets around a structural element (for example, bases flanking a hairpin loop or closing pair), so identical motifs can be scored differently depending on their local surroundings.
This hand-crafted, fine-grained context anticipated the learned features of modern deep models, which automatically capture similar local patterns and long-range dependencies via convolutions and attention.
This work fundamentally shifted the focus from meticulously measuring energy parameters to designing expressive, data-hungry statistical models, serving as a direct intellectual precursor to later deep learning methods.
While these early ML methods resulted in higher prediction accuracy, particularly on datasets structurally similar to their training data, they also introduced a significant risk of overfitting \citep{satoRecentTrendsRNA2023, szikszaiDeepLearningModels2022}.
This overfitting often manifested as a substantial drop in accuracy when these models were applied to predict secondary structures for RNA families that were not represented in their training data, thereby limiting their practical utility for newly discovered RNAs \citep{satoRecentTrendsRNA2023, szikszaiDeepLearningModels2022}
We will discuss the issue of overfitting and homology-aware benchmarks in more detail in the next section after a brief discussion on the history of the data used in the field.

\section{Datasets and generalization}\label{sec:datasets_and_generalization}

\subsection{The Evolution of Data}\label{subsec:datasets}

The history of data in RNA secondary structure prediction mirrors the field's methodological evolution, from physics-based models to data-driven paradigm.
This progression can be understood as a series of distinct eras, each defined by the nature and scale of the data that enabled its core computational approaches.

The first paradigm to establish itself was the \textbf{thermodynamic} one, which was defined not by datasets of examples, but by ``data as parameters'' \citep{andronescuComputationalApproachesRNA2010}.
The dominant minimum free energy (MFE) models, pioneered by Zuker and others, relied on a set of thermodynamic parameters that quantified the energetic cost or benefit of forming specific structural motifs like stacks and loops \citep{zukerOptimalComputerFolding1981, mathewsExpandedSequenceDependence1999}.
These parameters were not learned but meticulously measured through low-throughput optical melting experiments on short, synthetic RNAs \citep{mathewsIncorporatingChemicalModification2004}
This crucial information was compiled and disseminated through resources like the Nearest Neighbor Database (NNDB) \citep{turnerNNDBNearestNeighbor2010}, with the Turner rules becoming the \textit{de facto} standard \citep{mathewsExpandedSequenceDependence1999,mathewsIncorporatingChemicalModification2004}.
However, the accuracy of thermodynamic-based prediction was fundamentally bottlenecked by the precision of these physical measurements, and the model's simplifying assumptions highlighted the need for data derived from actual biological molecules \citep{andronescuComputationalApproachesRNA2010, szikszaiDeepLearningModels2022}.

This led to the development of \textbf{comparative and statistical paradigm}, which shifted the focus to curating ``gold standard'' biological structures.
These ground truths were sourced from high-resolution experiments like X-ray crystallography and, more scalably, from comparative sequence analysis, which identifies conserved pairings through co-varying mutations \citep{eddyRNASequenceAnalysis1994}.
Foundational databases like the Comparative RNA Web (CRW) Site \citep{cannoneComparativeRNAWeb2002} and the comprehensive RNA STRAND meta-database \citep{andronescuRNASTRANDRNA2008} aggregated thousands of these trusted structures, providing the first large-scale corpora for objective benchmarking.
This wealth of structural data enabled a powerful feedback loop, allowing the original thermodynamic parameters to be retrained and refined using biological examples \citep{andronescuComputationalApproachesRNA2010}.
Concurrently, it fueled the development of the first statistical predictors, such as CONTRAfold, which used conditional log-linear models trained on sequence-structure pairs derived from the Rfam database to learn scoring functions directly from data, demonstrating that a purely statistical approach could rival the accuracy of physics-based models \citep{doCONTRAfoldRNASecondary2006, kalvariRfam14Expanded2021}.

Concurrent with the curation of structural databases, a new data modality emerged from \textbf{chemical probing} experiments, which provide nucleotide-resolution information about the local structural environment of each base \textit{in vitro} and \textit{in vivo} \citep{strobelHighthroughputDeterminationRNA2018, weeksAdvancesRNAStructure2010, rouskinGenomewideProbingRNA2014}.
Rather than defining a complete structure, this data acts as a set of soft experimental constraints to guide computational predictions \citep{hajdinAccurateSHAPEdirectedRNA2013, reuterRNAstructureSoftwareRNA2010}.
Influential methods include SHAPE (Selective 2'-hydroxyl acylation analyzed by primer extension) \citep{merinoRNAStructureAnalysis2005}, which uses reagents to modify the ribose backbone at conformationally flexible nucleotides, and DMS (Dimethyl Sulfate) \citep{peattieChemicalProbesHigherorder1980}, which modifies the Watson-Crick face of unpaired adenine and cytosine bases.
The power of this approach lies in its direct integration with thermodynamic folding algorithms; the experimental reactivity scores are typically converted into soft energy constraints, thereby guiding the MFE search toward an experimentally supported conformation \citep{deiganAccurateSHAPEdirectedRNA2009, reuterRNAstructureSoftwareRNA2010, hajdinAccurateSHAPEdirectedRNA2013}.
The coupling of these techniques with next-generation sequencing created high-throughput methods like SHAPE-Seq and DMS-Seq, enabling transcriptome-wide structural interrogation and generating large-scale datasets of experimental constraints \citep{rouskinGenomewideProbingRNA2014, dingVivoGenomewideProfiling2014, mustoePervasiveRegulatoryFunctions2018}.

The advent of the \textbf{Deep Learning paradigm} created an unprecedented demand for data at a massive scale.
The millions of parameters in deep neural networks required far larger and more diverse datasets for effective training.
This need was met by the creation of key benchmarks like RNAStralign \citep{tanTurboFoldIIRNA2017} and ArchiveII \citep{samanbooyRNASecondaryStructure2022}, and most significantly, the bpRNA-1m \citep{danaeeBpRNALargescaleAutomated2018} database.
Aggregating over 100,000 structures, bpRNA-1m provided the necessary scale to train data-hungry models, and its standardized, non-redundant splits (TR0 for training, TS0 for testing) became the community standard for fair evaluation \citep{singhRNASecondaryStructure2019}.
However, this new power exposed a critical weakness: many models that performed well on TS0 failed to generalize to novel RNA families not seen during training, a problem that became known as the ``generalization crisis'' \citep{szikszaiDeepLearningModels2022}.
This realization mandated the development of more rigorous, homology-aware benchmarks.
Datasets like bpRNA-new \citep{satoRNASecondaryStructure2021}, composed of entirely new RNA families, and the stringent PDB-derived test sets (e.g., TS2), were created specifically to assess a model's ability to generalize beyond its training distribution \citep{satoRNASecondaryStructure2021, singhImprovedRNASecondary2021, frankeRNAformerSimpleEffective2024}.

Most recently, the \textbf{Foundation Model paradigm} entered the field, which leverages the vast, unlabeled sequence space of entire transcriptomes for self-supervised pre-training.
The central data source for this paradigm is RNAcentral \citep{thernacentralconsortiumRNAcentralHubInformation2019}, a meta-database containing tens of millions of non-coding RNA sequences.
By pre-training on this massive corpus, models like RNA-FM aim to learn the fundamental ``language'' of RNA without direct structural supervision \citep{chenInterpretableRNAFoundation2022}.
The frontier of scale is being pushed even further by efforts like the MARS database \citep{chenMARSRNAcmap3Master2024} and the Uni-RNA project \citep{wangUniRnaUniversalPreTrained2023}, which aim to aggregate over a billion nucleotide sequences from genomic and metagenomic sources.
This approach has created a two-tiered data ecosystem: massive, unlabeled sequence corpora are used for computationally intensive pre-training to build generalist models, while the smaller, high-quality labeled datasets like bpRNA are repurposed for the crucial tasks of fine-tuning and rigorous evaluation.

\subsection{The Generalization Crisis and the Mandate for Homology-Aware Benchmarking}\label{subsec:generalisation}
The challenge of generalizing predictions to novel RNA families has been a long-standing issue in machine learning for RNA structure prediction.
Early work on richly parameterized models like ContextFold had already demonstrated that while performance was high within known families, accuracy dropped considerably when tested on unseen ones, suggesting that models were learning family-specific features \citep{zakovRichParameterizationImproves2011, rivasRangeComplexProbabilistic2012}.

Despite these early insights, the initial wave of modern deep learning models, often reverted to less rigorous benchmarking based on simple sequence similarity cutoffs.
This led to reports of impressive but misleading accuracy \citep{szikszaiDeepLearningModels2022, justynaMachineLearningRNA2023}.
The subsequent "generalization crisis" was the widespread realization that these powerful new models were highly susceptible to overfitting and performed poorly when subjected to proper cross-family evaluation.
This was starkly investigated and quantified by \citet{szikszaiDeepLearningModels2022}, who demonstrated that a simple deep learning model's accuracy could plummet by 36\% in F1-score when moving from a flawed intra-family test to a rigorous inter-family one.

The response to this crisis was a community-wide re-adoption and formalization of rigorous, "homology-aware" benchmarking.
Building on the principles laid out by earlier studies, \citet{szikszaiDeepLearningModels2022} re-emphasized and standardized \textbf{family-fold cross-validation}—where entire RNA families are held out for testing—as the necessary gold standard for the deep learning era.
As discussed, several strategies have been investigated in order to help address this challenge beyond improved evaluation.
These include developing hybrid models like MXfold2 that ground deep learning in biophysical principles \citep{satoRNASecondaryStructure2021}, leveraging evolutionary information as in SPOT-RNA2 \citep{singhImprovedRNASecondary2021}, and designing models like RNAformer with meticulous homology-aware data pipelines from the ground up \citep{frankeRNAformerSimpleEffective2024}.
These approaches represent a concerted effort to build more generalizable and biologically faithful predictive tools.

\section{Deep Learning Methods}

\begin{figure*}[t]
\centering
\includegraphics[width=\linewidth]{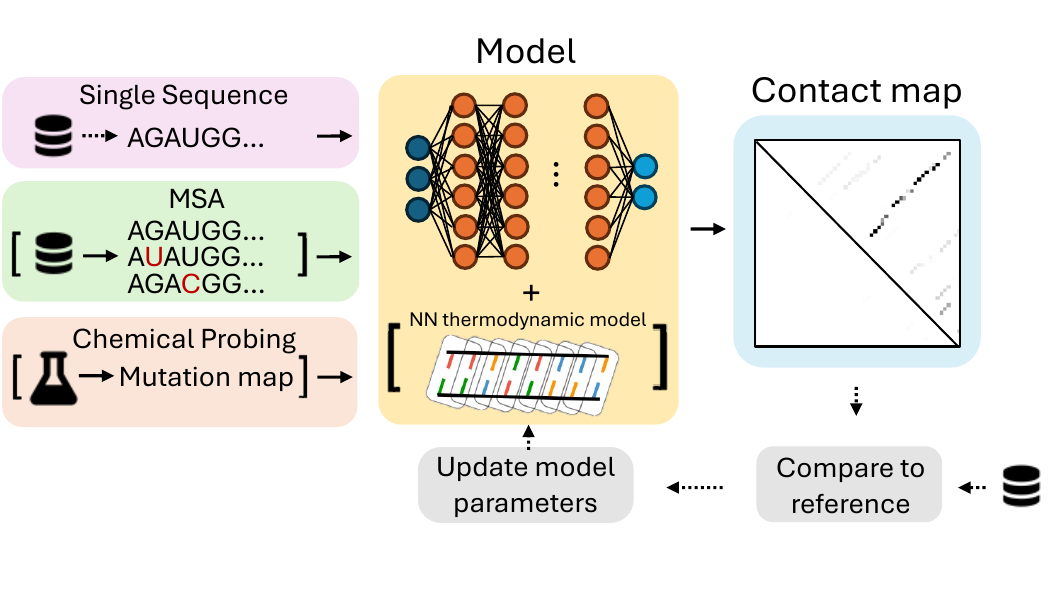}
\caption{
Schematic representation of deep learning methods for RNA secondary structure prediction (not including foundation models).
Dotted arrows indicate steps that are only included in training, and squared brackets indicate optional inputs.
Ab initio methods predict structure from a single RNA sequence only;
evolutionary methods leverage multiple sequence alignments (MSA) to capture co-evolutionary signals;
hybrid methods integrate deep learning with thermodynamic models or experimental data.
}
\label{fig:dl_methods}
\end{figure*}

The limitations of classical and early ML methods, together with the success of discriminative, richly parameterized models (e.g., CONTRAfold and ContextFold), set the stage for the current era dominated by deep learning.
Deep learning continues this data-driven line, moving from physics-based or statistically-tuned scoring functions to models that learn to predict a base-pairing contact map directly from the input sequence.

We can group these methods by the type of data they integrate.
The first category includes ab initio methods that predict structure from a single RNA sequence alone; the second leverages evolutionary information from multiple sequence alignments (MSAs); and the third consists of hybrid methods that combine deep learning with biophysical models or experimental data.
Figure \ref{fig:dl_methods} illustrates these workflows, from input data through training to final prediction.
We dedicate a separate subsection to Foundation Models (FMs), a rapidly evolving paradigm that pre-trains large neural networks on massive unlabeled RNA sequence datasets to learn the fundamental "language" of RNA, then fine-tunes them for specific tasks like secondary structure prediction, as shown in Figure \ref{fig:foundation_models}.

\subsection{Ab Initio (Single Sequence) Prediction}
These methods aim to predict the secondary structure using only a single RNA sequence as input, making them universally applicable and crucial for studying orphan RNAs.
A pioneering method, \textbf{SPOT-RNA} \citep{singhRNASecondaryStructure2019}, was directly inspired by the success of deep learning in protein contact map prediction. It applied the contact-map representation to RNA, modeling secondary structure as an adjacency matrix. The model utilized an ensemble of deep neural networks combining Residual Networks (Resnets) and 2D-Bidirectional LSTMs to predict this matrix, enabling it to model canonical, non-canonical, and pseudoknotted pairs. While it achieved a significant leap in performance on within-family test sets, subsequent independent studies revealed that it struggled to generalize to novel RNA families not seen during training \citep{szikszaiDeepLearningModels2022, justynaMachineLearningRNA2023}.
\textbf{UFold} \citep{fuUFoldFastAccurate2022} further advanced the image-based approach with a novel input representation that makes all potential interactions explicit. It converts the sequence into a multi-layered map where each of the 16 possible dinucleotide pairings is represented on its own L x L grid. This "image" is then processed by a U-Net architecture. This proved highly effective, showing substantial performance improvements and superior pseudoknot prediction on within-family datasets, though its ability to generalize to new families was also questioned in later studies \citep{szikszaiDeepLearningModels2022}.
\textbf{E2Efold} \citep{chenRNASecondaryStructure2020} integrated a Transformer model with an "unrolling algorithm," a technique that embeds hard structural constraints directly into the deep learning architecture. While it showed strong performance and high recall for pseudoknots on its original benchmarks, it was later found by multiple independent studies to be highly prone to overfitting, failing to generalize to new RNA families \citep{satoRNASecondaryStructure2021, fuUFoldFastAccurate2022}.
More recently, Transformer-based architectures, inspired by their transformative success in protein structure prediction with AlphaFold \citep{jumperHighlyAccurateProtein2021}, have gained prominence. \textbf{RNAformer} \citep{frankeRNAformerSimpleEffective2024} features a lean architecture with axial-attention blocks to efficiently capture long-range dependencies. Its main contribution is a novel homology-aware data pipeline that ensures a clean separation between training and test sets. This rigorous training and evaluation scheme was designed specifically to address the generalization crisis and allowed the model to achieve state-of-the-art performance on cross-family benchmarks, demonstrating strong generalization capabilities.
A novel generative approach is taken by \textbf{RNADiffFold} \citep{wangRNADiffFoldGenerativeRNA2025}, which uses a discrete diffusion model to progressively denoise an initially random contact map into a final, coherent structure. This process is guided by a conditional control component that fuses features from the sequence and, crucially, from pre-trained foundation models like RNA-FM (which we will discuss in Section \ref{sec:foundationmodels}). This method shows competitive performance across both within- and cross-family datasets and aims  at capturing dynamic, multi-conformational aspects of RNA structure.

\subsection{Evolutionary (MSA-based) Prediction}
To harness the powerful signal of co-evolution that proved so effective in classical comparative analysis, some of the most accurate deep learning models integrate information from MSAs.
\textbf{SPOT-RNA2} \citep{singhImprovedRNASecondary2021}, an evolution of its single-sequence predecessor, exemplifies this approach. It enriches its input features with evolutionary information by incorporating a Position Specific Score Matrix (PSSM) and a two-dimensional Direct Coupling Analysis (DCA) map, both derived from an MSA generated by its `RNAcmap` \citep{zhangRNAcmapFullyAutomatic2021} pipeline. This allows the model to learn from co-variation signals directly, showing a marked improvement over the original SPOT-RNA, particularly for complex interactions like non-canonical pairs. The method's accuracy scales directly with the number of available homologous sequences, achieving very high accuracy for RNAs with deep MSAs. However, this reliance on evolutionary data is also a key distinction; for orphan RNAs with very few homologous sequences, the original single-sequence SPOT-RNA can be more reliable. Furthermore, the method is computationally demanding and currently limited to sequences shorter than 1000 nucleotides due to its feature generation pipeline.
While primarily aimed at 3D structure prediction, the success of \textbf{trRosettaRNA} \citep{wangTrRosettaRNAAutomatedPrediction2023} further underscores the power of this strategy. Its pipeline begins with an initial secondary structure predicted by SPOT-RNA, which is fed into a transformer network along with an MSA. The network then predicts a comprehensive set of 1D and 2D geometric restraints (contacts, distances, and orientations) that are used to guide the final 3D folding. Notably, this process can correct inaccuracies in the initial secondary structure prediction, identifying interactions missed by SPOT-RNA and removing false positives. However, this corrective ability is a double-edged sword; in cases where the initial prediction is already highly accurate, potential conflicts between the different data sources can lead to a slight decrease in the final secondary structure's F1-score. Nevertheless, the high accuracy of its final 3D models is predicated on the high precision of its MSA-driven refinement and expansion of the initial 2D structural information.

\subsection{Biophysical (Hybrid) Approaches}
These methods seek the best of both worlds, combining the pattern-recognition strengths of deep learning with the rigorous framework of biophysical models. This is often done to improve generalization and ground the "black box" nature of deep learning in established physical principles.
One major strategy is to learn a data-driven scoring function that augments or replaces the standard thermodynamic energy parameters. \textbf{MXfold2} \citep{satoRNASecondaryStructure2021} is a prime example of this synergy. It employs a deep neural network to compute four types of folding scores which are then combined with Turner's free energy parameters. A key innovation is its use of "thermodynamic regularization" during training, which encourages the learned scores to remain close to the physical parameters, thereby preventing overfitting. This hybrid approach has proven highly robust in its authors' benchmarks, showing strong performance on datasets of unseen families where purely end-to-end models like E2Efold have struggled. However, other independent, rigorous cross-family evaluations have suggested that it, too, can struggle to generalize, indicating that integrating thermodynamic knowledge is a promising but not complete solution to the overfitting problem \citep{szikszaiDeepLearningModels2022}. Furthermore, its reliance on a Zuker-like DP algorithm restricts it to predicting non-pseudoknotted structures.
A second important hybrid strategy involves using machine learning to integrate diverse experimental data into the folding process. The model proposed by \textbf{Calonaci et al.} \citep{calonaciMachineLearningModel2020} is a sophisticated example. It uses a convolutional network to learn a mapping from 1D chemical probing data (e.g., SHAPE) and 2D co-evolutionary data (DCA) to pseudo-energy penalties. These learned penalties are then integrated directly into the RNAfold algorithm. A key advantage of this design is that the entire pipeline is differentiable, allowing the thermodynamic model to be part of the end-to-end training procedure, which was shown to significantly boost the population of the native structure in their tests.

Other methods achieve hybrid status through their algorithmic design or parameterization. It is worth noting that the core innovations of the following methods are not in machine learning themselves, but in their algorithmic or statistical frameworks. They are classified as hybrid because they are designed as flexible engines that can be parameterized by scores from either thermodynamic models or machine learning-based methods.
\textbf{LinearFold} \citep{huangLinearFoldLineartimeApproximate2019}, while primarily known for its linear-time complexity achieved via a beam search heuristic, can be parameterized with either traditional thermodynamic energies (LinearFold-V) or scores from machine learning models like CONTRAfold (LinearFold-C), making it a flexible hybrid tool.
Finally, some methods combine different scoring schemes at the ensemble level. \textbf{CentroidFold} \citep{satoCentroidFoldWebServer2009} operates on the principle of Maximum Expected Accuracy (MEA) rather than MFE. It predicts a "centroid" structure from a Boltzmann-weighted ensemble of possibilities by using a superior "g-centroid estimator." Its flexibility allows it to use parameters from Turner's model, CONTRAfold, or a combination, making it a statistical-mechanical hybrid that has demonstrated improved accuracy over pure MFE methods, though it is also limited to nested structures and has shown issues with generalization in cross-family tests \citep{szikszaiDeepLearningModels2022}.

\subsection{Foundation Models}\label{sec:foundationmodels}

\begin{figure*}[t]
\centering
\includegraphics[width=\linewidth]{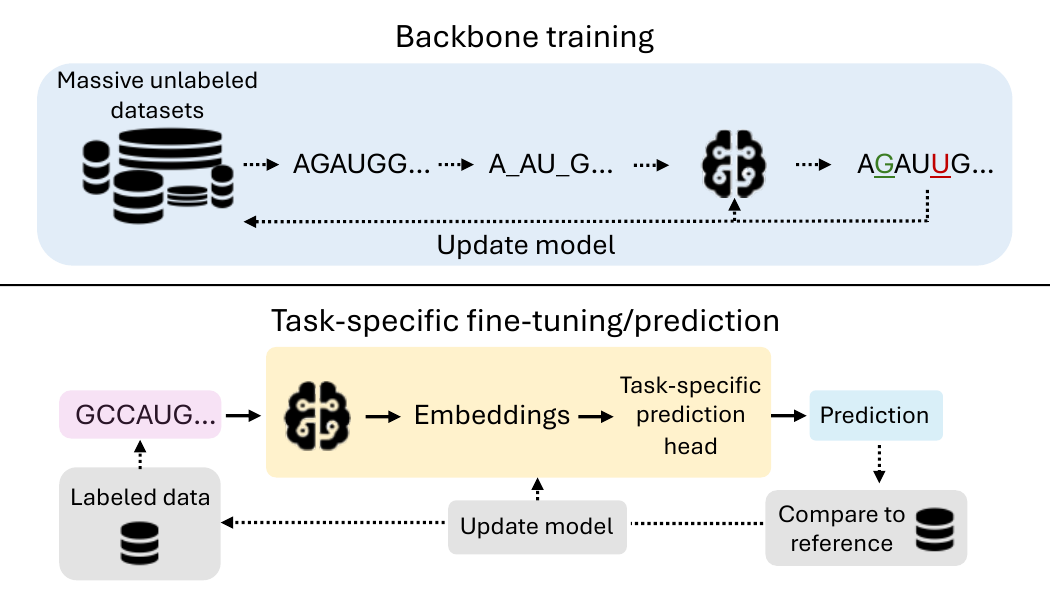}
\caption{
Schematic representation of backbone training (above) and task-specific fine-tuning/prediction (below) for RNA foundation models.
Dotted arrows indicate steps that are only included in training.
During backbone training, the model learns general "RNA language" features by predicting masked nucleotides from their surrounding context on massive unlabeled sequence datasets.
The pre-trained backbone can then be fine-tuned on smaller, labeled datasets for specific downstream tasks like secondary structure prediction.
}
\label{fig:foundation_models}
\end{figure*}
The generalization problem is inextricably linked to the "data bottleneck": the scarcity of diverse, high-quality experimental structures needed for supervised training.
Foundation Models (FMs), or RNA Language Models (LMs), represent a recent and rapidly evolving paradigm to address this.
Inspired by Large Language Models in natural language processing, this approach pre-trains large neural networks on millions of unlabeled RNA sequences to learn the fundamental "language" of RNA.
This paradigm aims to train large neural encoders to map sequences into rich, general-purpose embeddings that encode structural and functional features, which can then be fed into task-specific prediction heads, as illustrated in Figure \ref{fig:foundation_models}.

Scaling philosophies diverge: \textbf{UNI-RNA} \citep{wangUniRnaUniversalPreTrained2023} trains a 400M-parameter encoder on roughly one billion mixed genomic sequences, aiming for sheer quantity despite unresolved reproducibility concerns and critiques about heavy contamination from non-RNA fragments, whereas \textbf{AIDO.RNA} \citep{zouLargeScaleFoundationModel2024} keeps the corpus to 42M well-annotated RNAs, scales the model to 1.6B parameters, and releases checkpoints to emphasise quality-first curation.

Masked language modeling is the dominant training objective, exemplified by \textbf{RNA-FM} \citep{chenInterpretableRNAFoundation2022}, \textbf{ERNIE-RNA} \citep{yinERNIERNARNALanguage2025}, \textbf{RNAErnie} \citep{wangMultipurposeRNALanguage2024}, and \textbf{RiNALMo} \citep{penicRiNALMoGeneralPurposeRNA2025}, where nucleotides are masked during training to teach the model to predict them from context, thus learning sequence dependencies.
Additional objectives introduce structural supervision: \textbf{Orthrus} \citep{fradkinOrthrusEvolutionaryFunctional2024} augments masking with contrastive learning, where the model maximises similarity for biologically related pairs (orthologous, isoform pairs), while \textbf{ATOM-1} \citep{boydATOM1FoundationModel2023} and \textbf{RibonanzaNet} \citep{heRibonanzaDeepLearning2024} learn to regress chemical probing reactivities.
While these tasks add inductive bias, they rely on curated alignments or experimental measurements and therefore cover fewer sequences than masking objectives.
\textbf{MP-RNA} \citep{yangMPRNAUnleashingMultispecies2024} instead co-trains on ViennaRNA-derived pairing symbols, enabling structure-aware pre-training without experimental data, though at the risk of propagating biases from classical models.
\textbf{structRFM} \citep{zhuFullyopenStructureguidedRNA2025} extends this concept with a structure-guided masking loss that preferentially hides paired nucleotides so the backbone must internalize folding dependencies during pre-training.
\textbf{RNABERT} combines masked-token training with a structural alignment task that forces embeddings of homologous bases to be similar, thereby explicitly encoding structural conservation directly into the representation \citep{akiyamaInformativeRNABase2022}.

Architecturally, most RNA FMs are based on transformer encoders, from the foundational BERT-like RNA-FM \citep{chenInterpretableRNAFoundation2022} to RiNALMo \citep{penicRiNALMoGeneralPurposeRNA2025} that leverages modern LLM optimizations and pushes the scaling frontier with its 650M parameters.
Variants such as ERNIE-RNA add a physics-inspired pairwise bias to the first attention layer based on canonical pairing scores \citep{yinERNIERNARNALanguage2025},
\textbf{RNA-MSM} adapts the MSA transformer to operate on deep RNAcmap3-generated alignments \citep{chenMARSRNAcmap3Master2024} for Rfam families, overcoming the shallowness of manually curated MSAs \citep{zhangMultipleSequenceAlignmentbased2024}.
While these encoders enable rich context modeling, their quadratic complexity still limits practical input sizes, motivating designs like Orthrus, which leverages linear scaling of Mamba \citep{guMambaLinearTimeSequence2024} blocks, and hybrid schemes such as \textbf{HydraRNA} that alternate Hydra (bidirectional SSM) and attention layers to balance context range with efficiency \citep{fradkinOrthrusEvolutionaryFunctional2024, liHydraRNAHybridArchitecture2025}.

After pre-training, these models can be fine-tuned on smaller labeled datasets for specific downstream tasks like secondary structure prediction, often by adding lightweight prediction heads and optionally updating the backbone weights.

Homology-aware benchmarks underscore both the promise and limitations of these models, as \citet{zablockiComprehensiveBenchmarkingLarge2025} report a significant drop in performance in cross-family evaluations, with only ERNIE-RNA among the tested methods matching the robustness of thermodynamic-based algorithms.
When tested on the structures deposited in the PDB, LLMs performed worse than traditional methods and comparably to simple one-hot encodings \citep{zablockiComprehensiveBenchmarkingLarge2025}.
Despite these generalization challenges, foundation models obtained promising results in other independent benchmarks for specific tasks: the fine-tuned RibonanzaNet-SS \citep{heRibonanzaDeepLearning2024} ranked first on blind CASP15 secondary-structure assessments \citep{kryshtafovychCriticalAssessmentMethods2023} and ATOM-1 \citep{boydATOM1FoundationModel2023} surpassed larger MSA-dependent pipelines on structure prediction and achieved state-of-the-art results on stability benchmarks, yet comparative reviews continue to document robustness gaps on novel folds and long RNAs \citep{wangComparativeReviewRNA2025, zablockiComprehensiveBenchmarkingLarge2025}.
Current consensus emphasizes integrating evolutionary or experimental supervision earlier in pre-training and expanding homology-aware evaluations to bridge the remaining generalization divide \citep{wangComparativeReviewRNA2025, zablockiComprehensiveBenchmarkingLarge2025}.
The true robustness of all these models, however, will only become clear as they are more widely and independently benchmarked over time.

\section{Core Deep Learning Challenges: From Dynamics to Interpretability}\label{sec:challenges}
The integration of deep learning into RNA secondary structure prediction has brought significant gains but also new challenges concerning scientific validation, data scarcity, and the need for interpretable, biologically realistic models.

\subsection{Evolving Prediction Targets: From Static Blueprints to Dynamic Ensembles}

The central target of RNA structure prediction has evolved beyond the single \textbf{Minimum Free Energy (MFE)} structure.
This paradigm shift was driven by the recognition that the MFE is often misleading for functionally relevant RNAs, which typically exist as heterogeneous conformational ensembles where the most stable state may represent only a minor subpopulation.
This realization necessitated a move towards outputs that could capture this structural diversity.

Thermodynamic models offer a path towards this goal by using partition function algorithms to generate a Boltzmann-weighted sample of structures, approximating the full ensemble with predicted populations.
In principle, this is the ideal, full-information output.
However, its practical utility is fundamentally constrained by the limitations of the underlying energy model: its imperfect accuracy and its general inability to handle pseudoknots, non-canonical pairs, or the vast chemical diversity of modified nucleotides.

As a more practical, albeit lossy, summary of the ensemble, these models are often used to compute a 2D matrix.
This can be a \textbf{base-pairing probability matrix} (BPP), which has clear physical interpretability, or a more generalized \textbf{contact map}, which is the common output for deep learning models that trade physical meaning for the flexibility to represent any learned interaction \citep{fuUFoldFastAccurate2022}.
Both representations, however, obscure the correlational structure of the ensemble; the marginal probability of individual pairs does not capture the co-occurrence or mutual exclusivity of different structural elements, resulting in a significant information loss.

To refine predictions, a powerful strategy is to incorporate external data as \textbf{pseudo-energy terms} that perturb the energy landscape, a technique used in hybrid models like \textbf{ShapeKnots} or \textbf{MXfold2} \citep{hajdinAccurateSHAPEdirectedRNA2013, satoRNASecondaryStructure2021}.
Ultimately, the goal remains the accurate prediction of the full dynamic ensemble.
Single-molecule chemical probing (e.g., SHAPE-MaP \citep{smolaSelective2hydroxylAcylation2015} and DMS-MaPseq \citep{zubradtDMSMaPseqGenomewideTargeted2017}) records multiple modifications on individual molecules, preserving co-mutation patterns that reveal coexisting folds.
\textbf{DREEM} \citep{tomezskoDeterminationRNAStructural2020} clusters the reads obtained from each molecule to separate the ensemble-average profile into a small set of per-conformation reactivity profiles and their abundances; these profiles can then be used to constrain secondary-structure prediction.
\textbf{DRACO} \citep{morandiGenomescaleDeconvolutionRNA2021} scales this idea to longer RNAs via a windowed co-mutation graph, spectral model selection, and fuzzy clustering to determine and merge conformations.
\textbf{DANCE-MaP} \citep{olsonDiscoveryLargescaleCellstateresponsive2022} extends deconvolution by using a maximum likelihood clustering algorithm on MaP sequencing data to fit a Bernoulli mixture model.
This approach simultaneously extracts per-nucleotide reactivity, direct base pairing (PAIRs), tertiary interactions (RINGs), and populations for each conformational state.
By assigning individual reads to specific states, it enables state-specific correlation analyses that resolve structural features obscured in ensemble-average measurements.

\subsection{Emerging Frontiers and Persistent Hurdles}
Key challenges remain at the forefront of the field:
\paragraph{Chemical Complexity}
Most models operate on a simplified four-letter alphabet, ignoring the more than $150$ known post-transcriptional modifications observed \textit{in vivo}.
These modifications are not merely decorative; they directly alter base-pairing potential and are indispensable for stabilizing the complex tertiary folds essential for biological activity \citep{boccalettoMODOMICSDatabaseRNA2018}.
High-resolution structures (notably rRNAs and tRNAs) do contain numerous modifications, so structural data exist.
However, the space of modification chemistries is broad and for most individual modifications the available statistics are sparse, which hampers robust parameterization, benchmarking, and integration into folding models and ML features \citep{tanzerRNAModificationsStructure2019}.
Incorporating this chemical diversity remains crucial for biological realism and will require community resources aggregating per-modification thermodynamic and structural effects at scale.

\paragraph{Kilobase-Scale RNAs}
Accurately predicting the global architecture of long RNAs ($>1000$~nt) remains a major challenge, driven by both computational and biological complexity.
Computationally, many algorithms scale poorly with sequence length, making predictions intractable.
Biologically, the difficulty lies in capturing the correct hierarchy of local structures and the few crucial long-range interactions that define the global fold from a combinatorially vast search space \citep{szikszaiDeepLearningModels2022}.

\paragraph{Interpretability}
Deep learning models often function as ``black boxes,'' making it difficult to discern if they have learned generalizable biophysical rules or are simply fitting statistical patterns in the data.
This is a critical concern, as a lack of interpretability can be linked to poor generalization.
Hybrid approaches aim to address this by grounding the model in established physical principles.

\paragraph{Pseudoknots and Non-Canonical Pairs}
The accurate prediction of complex structural motifs remains a major hurdle.
Predicting pseudoknots within an energy minimization framework is an NP-complete problem, making exact solutions computationally intractable for all but the shortest sequences \citep{satoRNASecondaryStructure2021}.
Furthermore, the thermodynamic parameters governing their stability are poorly characterized.
Non-canonical base pairs, which are essential for stabilizing tertiary structure, are likewise excluded from most models due to a scarcity of experimental data.
While deep learning models are not bound by the same algorithmic constraints and can learn to predict these interactions, their accuracy remains limited, particularly for pseudoknotted base pairs where sensitivity is often low \citep{fuUFoldFastAccurate2022}.

\paragraph{Environmental Agnosticism}
Thermodynamic models, by incorporating experimentally measured enthalpy changes, can predict structures at different temperatures \citep{mathewsExpandedSequenceDependence1999}.
In contrast, most machine learning models are trained on sequence-structure pairs without environmental context.
They are therefore agnostic to physical parameters like temperature or ion concentration, limiting their ability to predict how an RNA's structure might change in different cellular or experimental conditions.
Also in this case, hybrid models that integrate learned scores with physical energy parameters, represent a promising step toward re-incorporating this biophysical realism.

\paragraph{Cellular Context and Ligand Binding}
RNA molecules fold \textit{in vivo} within a crowded cellular environment and their structures are often modulated by interactions with proteins, ions (e.g., $\mathrm{Mg}^{2+}$), and small-molecule ligands.
Riboswitches, for example, undergo functionally critical conformational changes upon ligand binding.
Most prediction methods, particularly single-sequence models, are blind to this context.
A key strategy to overcome this is the integration of experimental data from \textit{in vivo} chemical probing (e.g., SHAPE), which implicitly captures the effects of these cellular factors and can guide prediction algorithms toward more biologically relevant structures.

\paragraph{Standardized Prospective Benchmarking}
While data curation and evaluation practices have matured significantly, the field still lacks a community-wide, prospective benchmarking system analogous to the Critical Assessment of protein structure prediction (CASP) \citep{kryshtafovychCriticalAssessmentMethods2023}.
The success and rapid progress in protein folding, including the validation of AlphaFold, were driven in large part by CASP's role as an independent arbiter.
Establishing a regular, blind challenge for RNA secondary structure prediction could provide unbiased evaluation of true generalization capabilities, accelerate progress on persistent hurdles like pseudoknots and modified bases, and build community consensus on the genuine state-of-the-art.

\section{Discussion}

The field has shifted from principled yet constrained physics-based energy models to more flexible, data-driven deep-learning approaches.
This transition unlocked a new tier of predictive accuracy, yet simultaneously unveiled a profound challenge that has since reshaped the field's priorities: the generalization crisis.
The initial enthusiasm for deep learning's performance was rightly tempered by the discovery that many models were not learning the fundamental principles of RNA folding, but rather overfitting to family-specific features within the training data.
This rendered them unreliable for their most critical use case: the structural analysis of newly discovered or poorly understood RNAs.

This realization forced a necessary maturation within the community, compelling a move away from simplistic benchmarks toward rigorous, homology-aware validation standards.
The widespread adoption of family-based cross-validation has become the new gold standard, ensuring that modern methods are evaluated on their ability to generalize to unseen RNA families, not merely interpolate within known ones.
Strategies to address this challenge are now central to the field, including the development of hybrid models that ground learning in thermodynamic principles and the integration of co-evolutionary signals from multiple sequence alignments, which provide a powerful, albeit not universally available, source of structural constraint.

In response to both the generalization problem and the underlying scarcity of high-quality structural data, the current frontier is moving toward the use of foundation models.
By pre-training on millions of unlabeled RNA sequences, these models aim to learn the intrinsic "language" of RNA, capturing the statistical patterns that govern its structure and function without direct supervision.
This approach promises to create more robust and widely applicable predictors that are less dependent on the limited corpus of experimentally solved structures.
While still an emerging area, the development of these models represents a significant hope for breaking through the current data bottleneck.

Despite this progress, formidable challenges remain that will define the next era of research. The field must continue to evolve beyond predicting a single, static Minimum Free Energy structure; the crucial next step is to fully characterize the dynamic, conformational ensembles that define an RNA's functional landscape.
Moreover, the accurate prediction of complex motifs like pseudoknots and non-canonical pairs, which are often algorithmically intractable for physics-based models and a weak point for deep learning, remains a major barrier. Similarly, predicting the global architecture of long, kilobase-scale RNAs is largely unsolved, hampered by both computational complexity and the combinatorial explosion of possible long-range interactions. To achieve true biological realism, models must also incorporate the vast chemical diversity of post-transcriptional modifications and account for environmental context, such as ion concentrations, temperature, and in vivo ligand binding, all of which are critical for function but ignored by most current models. Addressing this suite of challenges will require not only new modeling strategies but also a community-wide, prospective benchmarking system, akin to CASP, to ensure unbiased validation and accelerate progress toward capturing the true, dynamic nature of RNA in the cell.

\section{Acknowledgements}
Guido Sanguinetti acknowledges co-funding from Next Generation EU, in the context of the National Recovery and Resilience Plan, Investment PE1 - Project FAIR “Future Artificial Intelligence Research”. This resource was co-financed by the Next Generation EU [DM 1555 del 11.10.22].

\bibliographystyle{unsrtnat}
\bibliography{bibliography}

\end{document}